\documentclass[11pt]{article}

\usepackage{amssymb}
\usepackage{amsmath}
\usepackage{amsfonts}
\usepackage{epsfig}
\usepackage{color}

\newcommand{\dslash}{\not{\hbox{\kern-2pt $\partial$}}}
\newcommand{\pslash}{\not{\hbox{\kern-2.3pt $p$}}}



\begin{document}
\newcommand{\be}{\begin{eqnarray}}
\newcommand{\ee}{\end{eqnarray}}
\newcommand\del{\partial}
\newcommand\nn{\nonumber}
\newcommand{\Tr}{{\rm Tr}}
\newcommand{\mat}{\left ( \begin{array}{cc}}
\newcommand{\emat}{\end{array} \right )}
\newcommand{\vect}{\left ( \begin{array}{c}}
\newcommand{\evect}{\end{array} \right )}
\newcommand{\col}{ \begin{array}{c}}
\newcommand{\ecol}{\end{array} }
\newcommand{\tr}{\rm Tr}
\newcommand\hatmu{\hat{\mu}}
\newcommand\noi{\noindent}
\newcommand{\bmini}{\begin{minipage}}
\newcommand{\emini}{\end{minipage}}

\newcommand{\bitem}{\begin{itemize}}
\newcommand{\eitem}{\end{itemize}}

\setlength{\baselineskip}{5.0mm}



\noindent{\bf \Huge
Handbook Article on\\[0.5cm] 
Applications of Random \\[0.5cm]
Matrix Theory to QCD}

\thispagestyle{empty}
\vspace*{2cm}

\noindent
{{\sc J.J.M. Verbaarschot}$^1$
\\~\\$^1$Department of Physics and Astronomy, 
Stony Brook University,\newline 
Stony Brook, NY 11794-3800}

\begin{center}
{\bf Abstract}
\end{center}
In this chapter of the Oxford Handbook of Random Matrix Theory
we introduce chiral Random Matrix Theories with the global
symmetries of QCD. In the microscopic domain, these
theories reproduce the mass and chemical potential dependence of QCD.
The main focus of this chapter is on the spectral properties 
of the QCD Dirac operator and relations between chiral Random
Matrix Theories and chiral Lagrangians.
Both  spectra of the anti-hermitian Dirac operator and spectra of the 
nonhermitian Dirac operator at nonzero chemical potential are discussed. 

\newpage

\tableofcontents
 \newpage
\section{Introduction}\label{intro}

Applications of Random Matrix Theory (RMT) to the physics of strong interactions
have a long history (see Chapter 2). RMT was introduced
to nuclear physics by Wigner 
to describe the level spacing distribution of nuclei
\cite{Wig55}. This paper inspired a large body of early work on RMT 
which is summarized in the book by Porter \cite{Por65}. An important
conceptual  discovery that emerged from this field is the large 
$N$ approximation. It first
appeared in the work of Wigner, and became an integral part of QCD through
the seminal work of 't Hooft \cite{tHo74}, which showed that the limit of
a large number of colors is
dominated by planar diagrams.
It was realized soon thereafter
that combinatorial factors can be obtained from matrix integrals
\cite{Bre78}. This culminated in the random matrix formulation
of quantum gravity in 2d, which is a sum over 
random surfaces that can~be~triangulated by planar diagrams.
(see Chapter 30 and \cite{DiF93} for a review).

The main focus of this chapter to apply  RMT to spectra
of the Dirac operator both at zero 
chemical potential, when the
Dirac operator is Hermitian, and at nonzero chemical potential, when the
Dirac operator is nonhermitian.  
Because the Euclidean Dirac operator
can be interpreted as a Hamiltonian, this application is closer in spirit
to the original ideas of Wigner than to the work of 't Hooft. However,
we have benefited greatly
from the mathematical techniques that were developed  for large $N$ QCD and 2d quantum
gravity.

Applications of RMT to QCD have been reviewed extensively in the literature
\cite{Guh97b,Jan98b,Ver00,Ver05,Ake07}. These papers offer both additional details and different points of view.
Phenomenological applications of RMT to QCD are
not discussed  in this chapter (see \cite{Ver00,Ake07} for reviews). 
 
In the first half of this chapter we introduce
chiral random matrix theory and its applications to QCD at zero chemical
potential. In the second half, chiral random matrix theories for QCD at nonzero chemical potential are discussed.
 
\subsection{Spontaneous Symmetry Breaking in RMT}
One of the essential features of Random Matrix Theory is spontaneous
symmetry breaking.  
The real part of the resolvent 
\be
G(z) = \frac 1N \left \langle {\rm Tr} \frac 1{z+D} \right \rangle,
\label{resolvent}
\ee
where the average is over the probability distribution of 
the ensemble of $N \times N$  anti-Hermitian random matrices, $D$, can be expressed 
as the replica limit
\be
{\rm Re}\, G(z) =  \lim_{n \to 0}\frac 1{2nN} \frac d{dz} \log(Z_n(z)), \qquad
{\rm with } \quad z \in {\mathcal R},
\ee
where
generating function is defined by
\be 
Z_n(z) = \langle {\det}^n(D+z){\det}^{n}(z-D) \rangle.
\label{zn}
\ee
For  $ z = 0 $, the generating function is invariant under $Gl(2n)$. This 
symmetry is broken spontaneously to $Gl(n) \times Gl(n)$ 
by a nonzero value of ${\rm Re}\, G(z)$
\be
\lim_{z\to 0} \lim_{N\to \infty} {\rm Re}\, G(z) \ne 0.
\ee
The order of the limits is essential -- the reverse order gives 
 zero.  This can be seen by expressing the resolvent in terms of
 eigenvalues of $D$. 

When we integrate the resolvent over the contour  $C$ 
in the complex plane that is the boundary of
$[-\frac \epsilon 2 < {\rm Re\, z} < \frac \epsilon 2 ]\times 
 [-\frac 12 \Delta x < {\rm Im\, } z < \frac 12 \Delta x]$ 
we obtain
\be
N \oint_C G(z) dz = 2\pi i N_{\Delta x}=2\pi i \rho(0) \Delta x,
\ee 
where $N_{\Delta x} $ is the number of eigenvalues enclosed by the contour
and $\rho(0)$ is the average spectral density around zero.
In the limit $\epsilon \to 0$ the l.h.s. is given by
$2i \Delta x {\rm Re}\, G(\frac \epsilon 2)$.
The discontinuity of the resolvent and $\rho(0)$ are thus
related by
\be
\lim_{\epsilon \to 0}{\rm Re}\,(G(\frac \epsilon 2)) = \frac {\pi \rho(0)}{N}.
\label{bc}
\ee
This formula is known as the Banks-Casher formula \cite{Ban80}.
It relates the order parameter for spontaneous symmetry breaking to the
spectrum of the associated operator.

When the spectrum of $D$ is reflection symmetric, i.e. $D$ and $-D$ have the 
same spectrum, then $\det (z-D) = \det (D+z)$, and the generating function
for the real part of the resolvent is given by
\be
Z_n(z) = \langle {\det}^n (D +z) \rangle.
\label{znd}
\ee 
This is the case when the random matrix ensemble has an involutive symmetry,
$
A D A = -D\quad  {\rm with} \quad A^2 =1,
$
which is the case for the Dirac operator in QCD.
                                                                                                                                                                                        
\subsection{The QCD Partition Function}
The Euclidean QCD partition function for $N_f$ quarks with mass $m_f$ is given by 
\be
Z_{\rm QCD} = \prod_{f=1}^{N_f}\langle {\det} (D+m_f) \rangle.
\ee
 The average is over gauge fields weighted by the Yang-Mills action. 
The gauge fields are elements of the Lie Algebra $SU(N_c)$ and  can be 
in the fundamental or adjoint representation. The theory that 
describes the strong interactions has $N_c =3$ with gauge fields in the
fundamental representation.
The
Dirac operator $D$ is a function of the ``random'' gauge fields. 

Applications of RMT to QCD differ in several respects from
other applications. 
First, the physical system itself 
is already a stochastic ensemble. Second, the QCD
partition function is not a quenched average, but the fermion determinant
describes the  quark degrees of freedom.
Third, the average is over Dirac operators with different rank. The reason
is that, according to the
Atiyah-Singer index theorem, the number of topological zero modes is 
equal to the topological charge of the gauge field
configuration. Because Dirac operators with different number 
of topological zero modes turn out to have spectra with different statistical
properties, we will treat each topological charge sector separately.
Fourth, QCD is a quantum field theory that has to be regularized 
and renormalized.
We notice
that the low-lying Dirac spectrum is gauge invariant and renormalizable
\cite{Giu09}.

\section{QCD and Chiral Random Matrix Theory}
In this section we construct a RMT  with the global
symmetries of QCD and determine the parameter range for which it is
equivalent to  the QCD.

\subsection{Symmetries of QCD}

To analyze the
global symmetries of QCD  we consider the Dirac operator for a finite
chiral basis. Then it is given by a matrix with the block structure 
\be
D =\mat 0 & iA \\ -iA^\dagger& 0 \emat,
\label{dblock}
\ee
where $A$ is a complex $N_+\times N_-$ matrix. 
Therefore, all nonzero eigenvalues occur in pairs $\pm \lambda_k$.
The number of zero eigenvalues is equal to $|N_+ - N_-|$ and is 
interpreted as the topological charge. 
Paired zeros may occur, but this a set of measure zero and of no interest.
 The block structure of (\ref{dblock}) is due to  the axial $U(1)$ symmetry. 
The partition function also has a vector $U(1)$ symmetry related
to the conservation of baryon charge.
For $N_f>1$ the Dirac operator is the direct sum of $N_f$ one-flavor
Dirac operators so that QCD
has the axial flavor symmetry
$U_A(N_f)$ and the vector flavor symmetry $U_V(N_f)$. 

For QCD with $N_c \ge 3$ and gauge fields in the fundamental representation
there are no other global symmetries. Because $SU(2)$ is pseudoreal, 
for $N_c=2$ the Dirac operator has an anti-unitary symmetry \cite{Ver94a}
\be
[UK,D] = 0
\ee
with $K$ the complex conjugation operator and $U$ a fixed unitary matrix
with $U^2 =1$. Then
it is always possible to find a basis for which $A$ becomes real \cite{Dys62},
and for $m=0$, we have that $\det D = \det^2 A$. For $N_f$ massless flavors the
quark determinant occurs in the partition function as 
${\det}^{2N_f} A$. This enlarges the  flavor symmetry group to $U(2N_f)$.
The third case is when gauge fields are in the adjoint representation. 
Then the Dirac operator also has an anti-unitary symmetry but now 
$U^2 = -1$ \cite{Ver94a}.
In this case it is possible to construct a basis for which 
the Dirac operator can be rearranged in 
self-dual quaternions \cite{Hal95b}.
The flavor symmetry is also enlarged to $U(2N_f)$.

QCD in three  Euclidean 
 space-time dimensions does not have an involutive symmetry. In that
case the flavor symmetry is $U(N_f)$ for gauge fields in the fundamental
representation. For two colors in the fundamental representation
and for any number of colors $\ge 2$ in the adjoint representation the
symmetry group is enlarged to $O(2N_f)$ \cite{Mag99a} 
and $Sp(2N_f)$ \cite{Mag99b}, 
respectively.

\subsection{Spontaneous Symmetry Breaking}
The features that mostly determine the physics of QCD at low energy
are spontaneous symmetry breaking and confinement. 
Because of confinement quarks and gluons do not appear in the 
physical spectrum so that 
QCD at low energy is
a theory of the weakly interacting Goldstone modes   associated with
the spontaneous breaking of chiral symmetry.

According to the Vafa-Witten theorem \cite{Vaf83}, global vector symmetries 
of vector-like gauge theories cannot be broken
spontaneously. The order parameter for  the breaking of the axial symmetry 
is the chiral condensate 
\be
\Sigma \equiv |\langle \bar \psi^a \psi^a \rangle| 
= \left |\lim_{m_a\to 0}\lim_{V\to \infty}\frac 1V  
\left \langle {\rm Tr } \frac 1{D+m_a}\right \rangle \right|
= \frac {\pi \rho(0)}V,
\label{bcqcd}
\ee
with $\rho(0)/V$ the spectral density of the Dirac operator per unit
volume of space time.
Because of the Banks-Casher relation, the absolute value of the chiral
condensate is flavor independent, but its sign is determined by the sign
of $m_a$.  

In Table \ref{table1} we give the symmetry breaking patterns \cite{Pes80,Vys85}
for the theories mentioned above.  We also give
the breaking pattern for QCD in three dimensions, but refer to the literature
for additional discussions \cite{Ver94b,Mag99a,Mag99b,Sza00,Dun02}.
  
\subsection{Chiral Random Matrix Theory}
 
\label{sec:chrmt}
Since the global symmetries of the Dirac operator are a  direct consequence
of its block structure and the reality properties of its matrix elements,
it should be clear how to construct a RMT with the same global 
symmetries: just replace the nonzero matrix elements by an ensemble of random
numbers.
Such chiral Random Matrix Theory (chRMT) is  defined by
\cite{Shu92,Ver94a}
\be
Z_{{\rm chRMT}}^{\beta,\nu}(\{m_k\}) = \int DA \prod_k \det \mat m_k& A  \\ A^\dagger & m_k \emat
P(A),
\ee
where $A$ is an $N_+\times N_-$ matrix, and the integration is over the
real (and imaginary) parts of $A$. 
 The reality classes are denoted by the Dyson index $\beta$ which is equal to
the number of degrees of freedom per matrix element and is the same as
in the corresponding QCD like theory (see Table \ref{table1}).
The properties of the chRMT partition function do not depend
on the details of the probability distribution. This is known
as universality \cite{Bre95,Jac96a,Ake96,Guh97a}
and justifies to simply average over a Gaussian distribution
\be
P(A) = c e^{-N\Sigma^2{\rm Tr}( A^\dagger A)} \qquad {\rm with }\quad
N = N_++N_-. 
\label{pa}
\ee

Both the Vafa-Witten theorem \cite{Vaf83} and the 
Banks-Casher formula  (\ref{bc}) apply to chRMT.
The global symmetry breaking pattern and the Goldstone manifold are therefore
the same as in QCD. 

In chRMT, $N$ is interpreted as the volume
of space-time. This corresponds to units where $N/V =1$ so that 
 $\Sigma$  can be written as $\Sigma V/N$ and
Eq. (\ref{pa})  becomes dimensionally correct. Notice that the matrix 
elements of the Dirac operator and its eigenvalues have the dimension
of mass. 
The normalization of (\ref{pa})
is such that $\Sigma$ can be interpreted as the chiral condensate that
satisfies the Banks-Casher relation.
\begin{table}[b!]
\begin{tabular}{|c|c|c|c|c|}
\hline
Theory & $\beta$ & Symmetry G &Broken to H& chRMT\\
\hline
\bmini{2.5cm}{\vspace*{0.2cm}Fundamental\\
$N_c \ge 3$, $d=4$ }\emini  & 2 &$U(N_f)\times U(N_f)$ & $U(N_f)$& chGUE \\[0.2cm]
\bmini{2.5cm}{\vspace*{0.2cm}Fundamental\\
$N_c = 2$, $d=4$ }\emini  & 1 &$U(2N_f)$ & $Sp(2N_f)$& chGOE \\[0.2cm]
\bmini{2.5cm}{\vspace*{0.2cm}Adjoint\\
$N_c \ge 2$, $d=4$\vspace*{0.2cm} }\emini  & 4 &$U(2N_f)$ & $O(2N_f)$& chGSE \\[0.2cm]
\hline
\bmini{2.5cm}{\vspace*{0.2cm}Fundamental  \\
$N_c \ge 3$, $d=3$ }\emini  & 2 &$U(N_f)$ & $U(N_f/2)\times U(N_f/2)$& GUE \\[0.2cm]
\bmini{2.5cm}{\vspace*{0.2cm}Fundamental\\
$N_c = 2$, $d=3$ }\emini  & 1 &$Sp(2N_f)$ & $Sp(N_f)\times Sp(N_f)$& GOE \\[0.2cm]
\bmini{2.5cm}{\vspace*{0.2cm}Adjoint\\
$N_c \ge 2$, $d=3$\vspace*{0.2cm} }\emini  & 4 &$O(2N_f)$ & $O(N_f)\times O(N_f)$& GSE \\[0.2cm]
\hline

\end{tabular}
\caption{Classification of QCD like theories in three and four dimensions. 
The Dyson index, $\beta$, is the number of degrees of freedom per matrix 
element ($\beta = 1$ and $\beta =4$ are interchanged 
for staggered fermions). 
}
\label{table1}
\end{table}

\subsection{Chiral Lagrangian}

The low-energy limit of QCD is given by a chiral Lagrangian and
necessarily has the name transformation properties as the QCD
partition function.  
The Lorentz invariant chiral Lagrangian 
to $O(M)$ and $O(p^2)$ is given by
\be
L = \frac 14 F^2 \,{\rm Tr} \,\del_\mu U \del_\mu U^\dagger - \frac 12 \Sigma
\,{\rm Tr }\, [MU^\dagger +M^\dagger U],
\label{chl}
\ee
where $U \in G/H$ with
$G$  the global symmetry group that is spontaneously broken to $H$ (see
Table \ref{table1}, and 
 $F$ is the pion decay constant.
In the domain 
\be 
M \ll \frac{\pi^2 F^2 }{\Sigma L^2} \ll \frac{ \pi^2 F^4}{\Sigma} 
\label{micro}
\ee
the kinetic term factorizes from the chiral Lagrangian \cite{Gas87}.
In this domain, known as the microscopic domain \cite{Ver94a},
the mass dependence  of the QCD partition function 
in the sector of topological charge $\nu$ is given by
\be
Z^\nu(M) = \int_{U\in G/H} dU {\det}^\nu U 
e^{-\frac 12 \Sigma {\rm Tr}[MU^\dagger +M^\dagger U]}.
\label{zm}
\ee
The first inequality in (\ref{micro}) can be rewritten as
$
1/{M_\pi} \gg L,
$
i.e. the pion Compton wavelength is much larger than the size of the box.
\begin{figure}[t!]
\unitlength1cm
\begin{center}
\begin{picture}(8,2.5)(2.2,3.5)
\centerline{\epsfig{figure=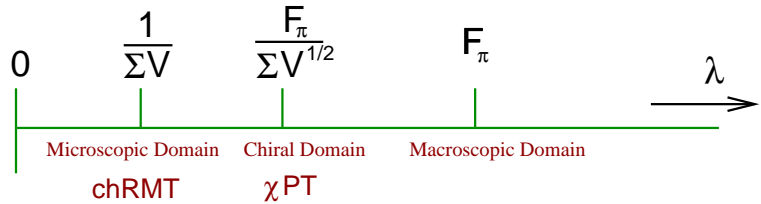,width=10cm,clip=}}
\end{picture}
\end{center}
\caption{Domains for the mass dependence of the QCD partition function.
\label{fig:scale}}
\end{figure}
The domain
where QCD is described by the chiral Lagrangian (\ref{chl}) including the
kinetic term, i.e. 
 where $M_\pi \sim p \ll \Lambda_{\rm QCD}$,  
will be called the chiral domain.
The zero momentum partition function (\ref{zm}) and   the chiral 
Lagrangian (\ref{chl}) are
the first term of the  $\epsilon$  and 
$p$ expansion \cite{Gas87}, respectively. Therefore these domains are also known
as the $\epsilon $ domain or $p$ domain corresponding to
a counting scheme where $M \sim 1/V$ and $M \sim 1/\sqrt V$,
in this order.

ChRMT can be reformulated identically in terms 
of a nonlinear $\sigma$-model even for finite size matrices \cite{Shu92,Jac96a}
(see also chapter 7). In this formulation the random matrix partition function
is given by a integral over
two types of modes, would be Goldstone modes with  mass
$\sim \sqrt {Nm}$ and massive modes with  mass $\sim \sqrt N$
(the number of integration variables does not depend on $N$). To leading
order in $1/N $ and $m$ the Goldstone
mode part factorizes from the partition function. 
Since the pattern of chiral symmetry breaking for QCD 
and chRMT is the
same, their mass dependence 
is the same in the microscopic domain (\ref{micro}).
 The correction terms are of $O(Nm^2)$, so that 
chRMT results are universal for $m\ll 1/\sqrt N$. With the
identification of $N$ as $V$, this corresponds to the $\epsilon$
domain.
 
In lattice QCD, and QCD in general, the topological charge
does not affect the block structure of the Dirac matrix but leads to 
nontrivial correlations between the matrix elements. 
Then it is natural
to work at fixed $\theta$-angle (in fact at $\theta = 0$) with partition
function given by $Z(m,\theta)$. 
However, properties of Dirac eigenvalues depend on the topological charge, and
in comparing lattice QCD and chRMT, Dirac spectra are sorted accordingly.
In chRMT,  topology is included by means
of the block structure of the Dirac matrix right from the start, 
and it is natural to work  with
fixed topological charge. The two partitions functions are related
by
\be
Z(m,\theta) = \sum_{\nu=-\infty}^\infty  e^{i\nu\theta} Z^\nu(m).
\ee
Since the spectrum of the Dirac operator depends on the topological charge
it makes sense to introduce the topological domain \cite{Leh09}, 
as the domain  where the average properties of the eigenvalues are sensitive
to the topological charge. This domain is expected to coincide with the
microscopic domain. 

\subsection{Generating Function for the Dirac Spectrum}
The generating function for the Dirac spectrum is also given by
(\ref{znd}) with the determinant of the physical quarks
contained in the average. Therefore, $z$ plays the role  of a quark
mass and the theory will have Goldstone bosons with squared mass
$ 
{2z \Sigma}/{F^2}. 
$
Therefore, for physical quark masses, i.e. quark masses that
remain fixed in the thermodynamic limit, we can choose
\cite{Ver95,Osb98b}
\be
z \ll \frac{F^2}{\Sigma L^2}\equiv E_{\rm Th}.
\label{thou}
\ee
In this domain the $z$-dependence of the generating function 
at fixed topological charge $\nu$ is
given by the zero momentum partition function (\ref{zm}) with 
quark masses equal to $z$.
The energy scale in (\ref{thou}) is known 
as the Thouless
energy.
A similar conclusion was reached in \cite{Jan98a}.
The volume dependence of the Thouless energy has been confirmed by
lattice simulations \cite{Ber98b,Ber99}.

The number of eigenvalues that is described by
chRMT scales as
$
E_{\rm Th}/ \Delta \lambda = {F^2L^2} /\pi.
$
Since $ F\sim \sqrt N_c$,
this number increases linearly
with $N_c$ \cite{Nar04}.

The determinants containing $z$ have to be quenched which can be done by the
replica trick (see chapter 8) or the supersymmetric method (see chapter 7). 
A supersymmetric version of
the chiral Lagrangian is known and the zero momentum integral has
been evaluated analytically \cite{Osb98a,Dam98b}.

\subsection{Chiral Random Matrix Theory and the Dirac Spectrum}

In an influential  paper that motivated the introduction of chRMT, \
Leutwyler and Smilga \cite{Leu92} proposed to expand the 
the QCD partition function at fixed topology
in powers of $m$ and equate the coefficients  to
the expansion of the same ratio for the low energy limit of QCD, i.e. for
the zero momentum partition function.
The sum rules are saturated
by eigenvalues in 
the microscopic domain which appear 
in the combination $\lambda_k V$.

With eigenvalues that scale as $1/V$,
the microscopic scaling limit of the spectral density can be defined as
\cite{Shu92,Ver94a}
\be
\rho_s(z) = \lim_{V\to \infty} \frac 1{V\Sigma} \rho\left ( \frac z{V\Sigma}
\right  )
\ee
with the microscopic scaling variable defined by
$
z = \lambda V \Sigma.
$
For $z \ll \sqrt V \Lambda^2$ the microscopic spectral density is
 given by chRMT. For all three values of the Dyson index, it can be
expressed in terms of the Bessel kernel 
\cite{For93}
\be
K_a(x,y) = \sqrt{xy} \frac {xJ_{a+1}(x)J_a(y)-yJ_a(x)J_{a+1}(y)}{x^2-y^2}.
\ee
The microscopic spectral density for $\beta = 2 $ is given by
\cite{Ver93}
\be
\rho_s^{\beta=2,a}(x) = \lim_{y\to x} K_a(x,y) =\frac 12 x( J^2_a(x) - J_{a+1}(x) J_{a-1}(x)),
\label{rho2}
\ee 
 with $ a= N_f + |\nu |$.
The microscopic spectral density for $\beta = 1$ and $\beta = 4$  can
be obtained by rewriting the partition function in terms of 
skew-orthogonal polynomials. For $\beta =1$
we find \cite{Ver93,For98,Ake08b}
\be 
\rho_s^{\beta =1,a}(z) &=& \frac 14 J_{a}(z) + \frac {z^a}4 \int_0^\infty dw w^{a} 
\frac{(z-w)}{|z-w|} 
\left (  \frac 1w \frac d{dw} - \frac 1z \frac d {dz} \right )
(zw)^{-a+3/2}K_{a-2}(w,z)\nn \\
&=&  \rho_s^{\beta =2,a}(z)+ \frac 12 J_a(|z|)\left ( 1 - \int_0^{|y|} dt J_a(t)
\right ) 
\ee
with $a= 2N_f+|\nu|$, and for $\beta =4$ 
the result is \cite{Nag95,Ake08b}
\be
\rho_s^{\beta=4,a}(z)&=& 2 z^2\int_0^1du  u^2 \int_0^1 dv(1-v^2)v^{-1/2}K_{a}(2uz,2uvz)\nn\\
&=&  \rho_s^{\beta =2,2a}(2z)- \frac 12 J_{2a}(2z)\int_0^{2|z|} dt J_{2a}(t).
\ee                        
where $a = N_f + 2|\nu|$.
Similar relations between the microscopic spectral density and the kernel
for $\beta = 2$ exist for an arbitrary invariant probability 
potential \cite{Sen98,Kle00}, and
can be exploited to show universality for $\beta =1$ and $\beta = 4$ from
the universality of the microscopic $\beta = 2$ kernel \cite{Ake96} (see
Chapter 6). 

It is of interest to
study the critical exponent of the spectral density at
a critical deformation of the Dirac operator for which $\rho(0) = 0$
which is a different universality class.
It is unlikely that  this critical exponent 
is equal to the mean field value of 1/3  \cite{Jac95}.
Other critical exponents can be obtained 
by fine tuning the probability distribution \cite{Ake97,Ake02a,Jan02}.

\begin{figure}[t] 
\unitlength1cm
\begin{center}
\begin{picture}(15.2,0.0)
\hspace*{-1cm}\centerline{\epsfig{figure=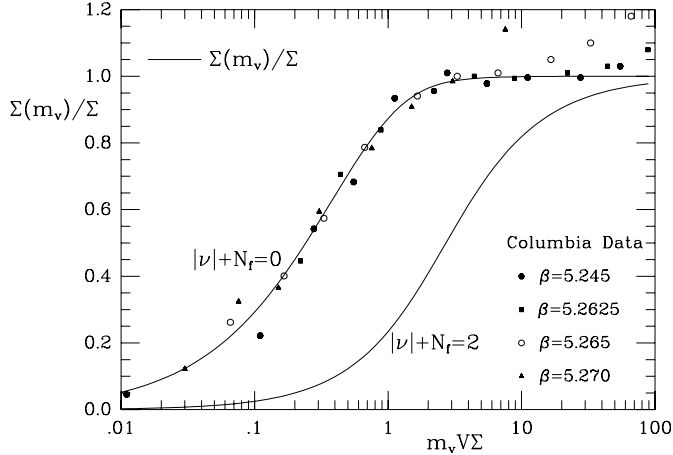,width=6cm,angle=-90,clip=}}
\end{picture}
\end{center}
\vspace*{6cm}
\caption{The valence quark mass dependence of the chiral condensate. Lattice data
obtained by the Columbia group \cite{Cha95} are compared to analytical chRMT
result given in Eq. (\ref{val}).
\label{fig:kyoto}}
\end{figure}

There are a large number of lattice results for the microscopic Dirac spectrum. 
The microscopic spectral density (\ref{rho2}) was first
observed for  lattice QCD Dirac spectra 
 through the mass dependence of the resolvent \cite{Ver95} defind as
 (earlier direct comparisons for the microscopic spectral
density were obtained for gauge field configurations given by a liquid
of instantons and anti-instantons \cite{Ver94c})
\be
\Sigma(m_v) = \left \langle \sum_k \frac 1{ \lambda_k + m_v} \right \rangle.
\label{sigmv}
\ee
The microscopic limit of $\Sigma(m_v)$, 
obtained by replacing the sum by an integral over the microscopic
spectral density, is given by a simple expression in terms of Bessel
functions \cite{Ver95}  
\be
\frac{\Sigma(m_v)}{\Sigma} = \hat m_v \left [ I_{N_f+|\nu|}(\hat m_v)K_{N_f+|\nu|}(\hat m_v)
+ I_{N_f+|\nu|+1}(\hat m_v)K_{N_f+|\nu|-1}(\hat m_v)\right ].\nn \\
\label{val}
\ee
In Fig. \ref{fig:kyoto} we compare this result to 
the chiral condensate (\ref{sigmv}) obtained
from  lattice simulations 
\cite{Cha94,Cha95} for
two flavors and various
values of the coupling constant. 
The average is over the Yang-Mills action and the fermion determinant
with the masses in the 
fermion determinant  kept fixed.
Agreement is found 
with the quenched result because the
physical quark masses are much larger than the microscopic scale. 
There is no  dependence
on the topological charge because the staggered lattice 
fermions are not close enough to the continuum limit
(This point was investigated in more detail in \cite{Dam99c,Far99}. More
recently it was shown \cite{Won04,Fol04,Fol05} that dependence on
topology as predicted by chRMT is reproduced by staggered fermions
if we are sufficiently close to the continuum limit). 
 Similar lattice results have been obtained for $\beta =1 $
and $\beta =4$ \cite{Dam99b} together with the corresponding analytical 
results.

 \begin{figure}[t!]
\unitlength1cm
\begin{center}
 \centerline{
\epsfig{figure=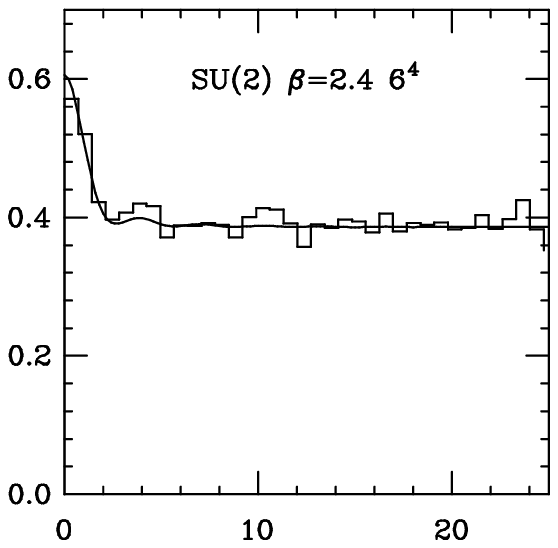,width=3.8cm,bb=250 315 425 500,clip=  } \vspace*{-1cm}
\epsfig{figure=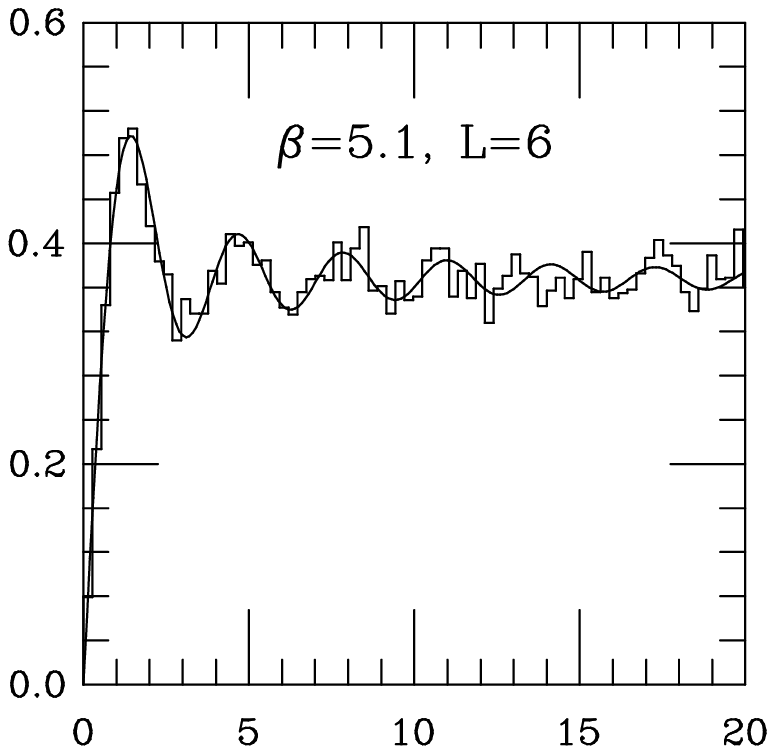,clip=,width=4.5cm}
\epsfig{figure=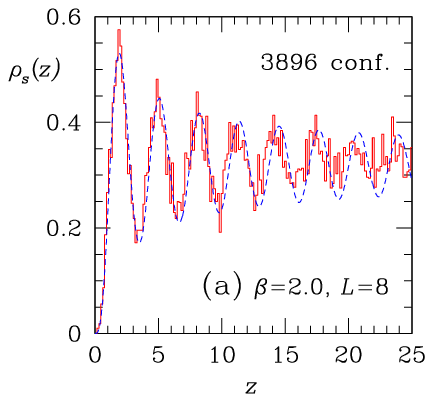,width=4.0cm,clip=} 
}
\end{center}
\caption{
Microscopic spectral density for $SU(2)$ gauge group in the adjoint
representation  
(left, taken from \cite{Edw99})
and in the fundamental  representation (right, taken from \cite{Ber98a}), and for
$SU(3)$ gauge group in the fundamental representation (middle, taken from \cite{Dam98a}). 
\label{fig:micro}}
\end{figure}

The first direct observation of the micropscopic spectral density in QCD
was made for staggered fermions with
two colors in the fundamental representation \cite{Ber97a} 
for which the Dyson index
is $\beta = 4$ (see right panel of Fig.~\ref{fig:micro}). Results for
QCD with $N_c=3$ were obtained in \cite{Dam98a,Goc98} 
(Fig.~\ref{fig:micro}, middle). The left panel of Fig.~\ref{fig:micro} is for 
staggered fermions in the adjoint representation \cite{Edw99}.

The distribution of the smallest Dirac eigenvalue 
is given by  $P_{\rm min}^{\beta\,\nu}(s) = - E'(s)$  with $E(s)$ 
defined as  the probability that there
are no eigenvalues in the interval $[0,s \rangle$.
In Table~\ref{table2} we summarize analytical results for the quenched
case \cite{Ede88,For93,Wil97,Dam00}. The result for $\nu = 0$ is particularly
simple.
Expressions for the $k$'th smallest eigenvalue at arbitrary quark mass, 
topological charge and Dyson index  are known as well \cite{Wil97,Nis98,Dam00}. A useful measure to compare lattice QCD results with chRMT predictions
is the ratio of low lying eigenvalues \cite{Giu03,Fod09}. 
Agreement with the topology and mass dependence has become an important
tool in lattice QCD to test the lattice implementation of chiral symmetry
and topology. 

The joint probability distribution of chRMT only 
depends  on $N_f$ and $\nu$ through  the
combination $2N_f +\beta \nu$. This property, known as flavor-topology
duality \cite{Ver97},  has been  observed
in lattice QCD Dirac spectra  \cite{Fuk07}.

Contrary to correlations of low-lying eigenvalues, 
bulk spectral correlations can be investigated by spectral averaging, and
for large lattice volumes,
the Dirac spectrum of a single gauge field configuration is
sufficient to obtain statistically significant correlators.
Excellent agreement with the 
Wigner-Dyson ensembles was obtained \cite{Hal95a}
without sign of a Thouless energy. 
It turns out \cite{Guh98} that  
the  Thouless energy scale is due to ensemble averaging.
The conclusions is that there is no spectral ergodicity 
beyond the Thouless energy.
\begin{table}[h!]
\begin{tabular}{|c|c|c|c|}
\hline
$e^{\beta\zeta^2/8}P_{\rm min}^{\beta, \nu}(\zeta)$& $\nu = 0$ & $\nu =1$ & 
\bmini{3cm}general $\nu$\\ $\nu$ odd for $\beta =1$ \emini
 \\
\hline
$\beta = 1$ &$ \frac 14(2+\zeta)e^{ -\zeta/2}$ 
& $\zeta  I_3(\zeta)$&
          $\zeta^{(3-\nu)/2} {\rm Pf}[(i-j)I_{i+j+3}(\zeta)]$ \\
$\beta = 2$ & $\zeta $ & $\frac \zeta 2  I_2(\zeta)$&
 $\frac \zeta 2 \det I_{i-j+2}(\zeta)$\\
$\beta = 4 $&$ \frac 12 (e^\zeta(\zeta-1) +e^{-\zeta}(\zeta+1))$&
& $\zeta^{4\nu+3}  (1+\sum_ja_j(|\nu|) \zeta^j $) \\
\hline
\end{tabular}
\caption{Results for the distribution of the smallest Dirac eigenvalue for
the quenched case. In the last row $a_j$ is an expression in terms 
of a sum over partitions of $j$. For explicit expressions
we refer to \cite{Ber98a}.}
\label{table2}
\end{table}
\vspace*{-0.5cm}
\subsection{Integrability}
For $\beta = 2$ the partition function of invariant random matrix theories 
can be interpreted 
as a partition function of noninteracting fermions which is 
an integrable system. This is the reason that the low energy QCD
partition function, the unitary matrix integral (\ref{zm}), 
obeys a large number
of remarkable relations. 
The $N_f$ flavor partition function in the sector of topological charge
$\nu$ can be written as \cite{Bro81a,Bro81b,Guh96,Jac96b}
\be
Z^\nu_{N_f}(m_1, \cdots, m_{N_f}) = \frac {\det [x_k^{k-1} I_\nu^{(l-1)}(x_k)]_{k,l =1,\cdots, N_f}}
{\Delta(\{x_k^2\})}, \quad x_k = m_k V \Sigma.
\label{zdiffmass}
\ee
In the limit $x_k \to x$ this partition function  reduces to
a Hankel determinant
\be 
Z^\nu_{N_f} = \det [ (x\del_x)^{k+l}I_\nu(x)]_{k,l=0,\cdots,N_f-1}.
\ee
Applying the Sylvester identity \cite{For02} relating the determinant of a matrix to  co-factors 
gives the Toda lattice equation \cite{Kan02,Spl03a}
\be
(x\del_x)^2 \log Z_{N_f}^\nu(x) = 2N_f x^2 \frac {Z^\nu_{N_f+1}(x) Z_{N_f-1}^\nu(x)}
{[Z_{N_f}^\nu(x)]^2}.
\ee
After taking the replica limit of this recursion relation we arrive at 
the following compact
expression for the resolvent \cite{Spl03a}
\be
x\del_x x G(x) = \lim_{N_f \to 0} \frac 1{N_f} \del_x \log Z_{N_f}^\nu(x)
=
2x^2 Z_1^\nu(x) Z_{-1}^\nu(x).
\ee
This factorized form is a general property of the spectral density and correlation
functions of RMT's with $\beta = 2$.
Application of the replica limit to a discrete recursion relation does
not require analyticity in the replica variable and this way  problems
with the replica limit can be  circumvented \cite{Ver85,Kan02} (see 
Chapter 8). 

As a consequence of integrability relations, the zero momentum partition
function satisfies 
Virasoro constraints. They provide efficient way to 
determine the coefficients of the small
mass expansion \cite{Dam99a,Dal01}.

In  addition to the relations discussed above  
we would like to mention the following relations:
i) The Toda lattice equation
can be formulated for finite size random matrices by exploiting the 
properties of orthogonal polynomials \cite{Ake04}.
ii) 
In the microscopic domain  the 
partition functions in 3 and 4 dimensions are related by \cite{Ake99,Ake00b,And04}
\be
Z_{{\rm QCD}_3}^{2N_f}(\{x_k\}) = Z^{\nu=-1/2}_{N_f}(\{x_k\}) Z^{\nu=1/2}_{N_f}(\{x_k\}).
\ee
iii) $k$-point spectral correlation functions can be expressed 
into partition functions
with $\beta k$  additional flavors \cite{Ake98}.
iv)
The correlation functions of invariant RMTs can be expressed in terms
of the two-point kernel. This leads to consistency relations between various partition functions \cite{Ake98}
v) Interpreting the quark mass as an additional eigenvalue leads to
relations between correlators of
massive and massless partition functions \cite{Ake00a}.

\section{ChRMT at Nonzero Chemical Potential}

An important application of chRMT is to QCD at nonzero chemical 
potential $\mu$.
In that case the  Dirac operator 
is given by
\be
D(\mu)= D(\mu =0) + \mu \gamma_0.
\ee
Since $D(\mu=0) $ is anithermitian, $D(\mu \ne 0) $ 
has no hermiticity properties, and its eigenvalues are scattered in the complex plane. 
Because the determinant of the Dirac operator is complex, the QCD partition function at
$\mu \ne 0$ is the average of a complex weight, and unless
the chemical potential is small, it cannot be simulated by Monte-Carlo
methods. For that reason chRMT has been particular helpful to answer 
questions that could not be addressed otherwise. In particular, the 
following issues have been clarified:
i) The nature of the quenched approximation \cite{Ste96b}.
ii) The relation between the chiral condensate and the spectrum
 of the Dirac operator for QCD with dynamical quarks \cite{Osb05}.
iii) The expectation value of the phase of the fermion determinant \cite{Spl06}.
iv) The low-energy limit of phase-quenched QCD and the spectrum of
the Dirac operator \cite{Tou00}.
v) The geometry of the support of the Dirac spectrum \cite{Tou00,Osb08b}.

The QCD partition function at nonzero chemical potential is given
by 
\be
Z_{\rm QCD} = \langle \prod_{k=1}^{N_f} \det(D+m_k + \mu_k \gamma_0)\rangle.
\ee
where the average is over the Yang-Mills action.
The chemical potential for different flavors is general different.
Two important special cases are: $\mu_k = \mu$ when$\mu$ is the baryon chemical potential, and the case for an even number of flavors
$N_f = 2n$  with  $\mu_k = \mu$ for $k= 1,\cdots, n$ and
$ \mu_k = - \mu$  for $k = n+1, \cdots, 2n$. 
In the second case the partition function is positive definite
because
\be 
 \det (D+m - \mu\gamma_0)     
=\det (D^\dagger +m + \mu\gamma_0)
=  {\det}^* (D+m +\mu\gamma_0).
\ee
For $n=1 $, $ \mu$  can be interpreted as an isospin 
chemical potential \cite{Son00}.
Since the determinant appears together with its complex conjugate, this
partition function is also known as the phase quenched two-flavor
partition function.

\subsection{Dirac Spectrum}

A particular useful tool for studying
the spectrum of a nonhermitian operator is the resolvent
(\ref{resolvent}).
The spectral density is given by
\be
\rho(z,z^*) = \langle \sum_k\delta^2(z-\lambda_k) \rangle=\frac 1\pi \frac d {dz^*} G(z).
\ee
and 
can be interpreted as the two-dimensional electric field at $z$ of charges
located at $\lambda_k$. When the eigenvalues are on the imaginary axis 
this picture illustrates that $G(z)$ has a discontinuity when $z$ crosses
the imaginary axis. When eigenvalues are not constrained by Hermiticity,
because of level repulsion, they will scatter into the complex plane. 
Using the electrostatic analogy, the
resolvent will be continuous. If $z$ is outside the spectrum,
 $G(z)$ is   analytic in $z$.
The resolvent cannot be expressed in terms of Eq. (\ref{znd}) but rather as
\be
G(z) = \lim_{n \to 0} \frac 1n \frac d {dz} Z_n(z,z^*),
\ee
where $Z_n(z,z^*)$ is  the phase quenched partition function \cite{Ste96b}
\be
Z_n(z,z^*) = \langle {\det}^n (D +z) {\det}^n(D^\dagger+z^*) \rangle.
\label{znmu}
\ee

 Lattice QCD Dirac spectra were first calculated in \cite{Bar86}.
As remarkable features we note that the spectrum is approximately
homogeneous, and that it has a sharp edge
which both are explained by chRMT.

\subsection{Low-Energy Limit of QCD and Phase Quenched QCD}

According to the definition of the grand canonical partition function,
 the free energy  at low temperature does not depend on the
chemical potential  until it is equal to the  lightest physical
excitation (per unit charge) with charge conjugate to $\mu$. 
For QCD this  implies that the chiral condensate at zero temperature 
does not depend on $\mu$ until $\mu = m_N/3$ (with $m_N$ the nucleon mass). 
The Dirac spectrum, is $\mu$ dependent, though, which seems to violate
the Banks-Casher relation \cite{Ban80}. This problem is known as
the 'Silver Blaze Problem'  \cite{Coh03}.

At nonzero isospin chemical, $\mu_I$,
the critical chemical potential is equal to
 $\mu_I = m_\pi/2$. 
Beyond this point,  pions will Bose condense. For light
quarks, this phase transition can be studied by chiral perturbation 
theory, and for quark masses in the microscopic domain it is
described by chRMT. At nonzero  $\mu_I$,
the 'Silver Blaze Problem' is that at zero temperature the chiral condensate remains
constant until $\mu_I = m_\pi/2$, while the spectral density depends on $\mu_I$.
The solution is easy: according to
the electrostatic
analogy, the 
'electric field', i.e. the chiral condensate, is constant outside 
a homogeneously charged strip. This implies that the
width of the strip is determined by the relation $\mu_I = m_\pi/2$
\cite{Gib86,Tou00}.
Indeed, in terms of eigenvalues, the critical point is when
the quark mass hits the boundary of the spectrum.

\subsection{Chiral Lagrangian at Nonzero Chemical Potential}

Chiral symmetry remains broken at small  nonzero 
chemical potential. Therefore, also in this case, 
the low-energy limit of QCD 
is given by a theory of weakly interacting
Goldstone bosons. As is the case at zero chemical potential, 
the $U_L(N_f)\times U_R(N_f)$ invariance of the 
partition function is broken spontaneously to $U_V(N_f)$. 
The invariance properties 
of QCD should also hold for the Lagrangian
that describes the low-energy limit of QCD. In particular, because
the chemical potential is an external vector potential, it only enters in 
the combination of the covariant derivative \cite{Kog99}
\be
\nabla_\nu U = \del_\nu U - [B_\nu, U], \qquad {\rm with}\quad
B_\nu = {\rm diag}(\{ \mu_k\})\delta_{\nu,0}.
\ee
Together with the mass term, the $O(p^2)$ chiral Lagrangian is thus
given by
\be
{\cal L} = \frac {F^2}4 \nabla_\nu U \nabla_\nu U^\dagger
-\frac 12 \Sigma {\rm Tr}( MU + MU^\dagger).
\label{lmu}
\ee
Eq. (\ref{lmu}) shows that the chiral Lagrangian is determined by two constants.
Since the Dirac spectrum at $\mu \ne 0 $  is also determined by two constants,
the eigenvalue density and the width of the spectrum, we can extract the
low energy constants from the geometry of Dirac spectrum.

It has been argued that at sufficient large chemical potential QCD
will be in a color-flavor locked phase with spontaneously  
broken color-flavor symmetry.
In this phase the chiral
condensate vanishes, and  next order terms in the chiral
expansion, which are quadratic in the quark mass, have 
to be taken into account. Universal results are obtained 
by scaling the Dirac eigenvalues with $\sqrt V$ and Leutwyler-Smilga
sum rules have been derived both for QCD with three colors \cite{Yam09}
and QCD with two colors \cite{Kan09}.

\subsection{Chiral Random Matrix Theories at $\mu \ne 0$}

In a suitably normalized chiral basis the Dirac operator at nonzero
chemical potential has the block structure
\be
D(\mu) = \mat 0 & id +\mu \\id^\dagger +\mu &0 \emat.
\label{diracmu}
\ee
A chiral random model at nonzero chemical potential is obtained
\cite{Ste96b} by replacing the matrix elements of $d$ and $d^\dagger$ 
by an ensemble of random numbers exactly as 
in section \ref{sec:chrmt}. Also in this case, 
because of expected universality \cite{Ake02b}, it is justified to
simplify the model by choosing a Gaussian distribution.

As is the case for $\mu = 0$, 
we can distinguish three different nonhermitian chRMTs
 \cite{Hal97},
with complex matrix elements ($\beta = 2$), with real matrix 
elements ($\beta = 1$), and with self-dual quaternion matrix elements
($\beta =4$). 
They apply to the same cases as discussed
in table \ref{table1}. The full classification of nonhermitian ensembles
is based on the  Cartan classification of symmetric spaces
\cite{Zir96,Ber01,Mag07}.

 The random matrix model (\ref{diracmu}) is not unique. Adding $\mu$ in
a different way results in the same chiral Lagrangian as long as the
invariance properties of the matrix model remain the same. One drawback
of the model (\ref{diracmu}) is that the overall unitary invariance has been
lost so that methods that rely on the joint probability distribution 
of eigenvalues cannot be used.
 A model that does have a representation in terms of eigenvalues is defined
by \cite{Osb04}
\be
D = \mat 0 & id +\mu C \\ id^\dagger +\mu C^\dagger & 0 \emat,
\label{diracjames}
\ee
where $C$ and $d$ are complex random matrices 
with the same distribution.

A rerun of the  arguments of \cite{Gas87,Dam06} 
in the microscopic domain 
\be
m^2 V\ll 1 \qquad   {\rm and}\qquad  \mu^4 V \ll 1,
\ee
shows that the partition
function corresponding to  the chiral Lagrangian (\ref{lmu})
factorizes into 
a zero momentum part and a nonzero momentum part. 
The chemical potential and mass dependence reside in the zero momentum
part.

In the microscopic  domain, invariance properties of QCD at $\mu \ne 0$ 
that rely on global symmetries can also hold for
chRMT at $\mu \ne 0$ and  give the same invariant terms in the zero momentum sector. Therefore, in this  domain, the QCD partition function
is given by chRMT at $\mu \ne 0$. Mean field studies only involve the
 zero momentum part of the 
chiral Lagrangian. Therefore, mean field results \cite{Kog00}
can also be derived from chRMT.

Applying the above arguments to the generating function for the Dirac spectrum
we obtain the zero momentum partition function
 \be
Z_n^\nu(z,z^*;\mu) &=& c \int_{U \in U(2n)} dU {\det}^\nu U e^{-\frac {VF^2\mu^2}4
{\rm Tr} [U,B][U^\dagger,B] + \frac 12 \Sigma V {\rm Tr}(MU + MU^\dagger)}\nn \\
&& {\rm  with}\quad
B = \Sigma_3 \qquad {\rm and} \qquad M = \mat z& 0 \\ 0 & z^* \emat.
\label{zeff}
\ee

At the mean field level the resolvent is independent of the replica index 
and the partition function can be analyzed for $n=1$. For the resolvent
we find,
\be 
G(z) = \Sigma  \qquad {\rm and}\quad  G(z) = \frac{\Sigma^2 (z+z^*)}{4\mu^2F^2}
\ee
for ${\rm Re}\, z > 2\mu^2 F^2 /\Sigma$ and 
${\rm Re}\, z < 2\mu^2 F^2 /\Sigma$, respectively. 
The eigenvalues are therefore distributed homogeneously inside
a strip with width $ 4F^2 \mu^2 /\Sigma$. In agreement with lattice
simulations \cite{Bar86}, the eigenvalue density has a sharp edge whereas
the resolvent is continuous at this point. If $z$ and $z^*$ are interpreted as
quark masses,  the squared mass of the corresponding Goldstone bosons is equal to
$
m_G^2 = {(z+ z^*)\Sigma}/{F^2}.
$
The condition ${\rm Re}\, z < 2\mu^2 F^2 /\Sigma$
can then be written as $\mu = m_G/2$, in agreement 
with physical considerations.

\subsection{Integrability of the Partition Function}
\label{sec:integrability}

Remarkably, as was the case $\mu = 0$,  
the zero momentum partition function (\ref{zeff})
can be rewritten  in terms  of a Hankel like determinant \cite{Spl03a,Spl03b}
\be
Z_n^\nu(z,z^*,\mu ) = D_n (z z^*)^{n(1-n)} \det[(z\del_z)^k (z^*\del_{z^*})^l 
Z_1^\nu(z,z^*,\mu)
]_{k,l=0,1,\cdots,n-1}.\hspace*{0.5cm}
\label{hankel-form}
\ee
This form  responsible for the integrable
structure of the partition function (\ref{zeff}). Most notably, it satisfies the 
Toda lattice equation 
\be
z\del_z z^*\del_{z^*} \log Z_{n}^\nu(z,z^*,\mu) =
\frac{\pi n}2 (zz^*)^2 \frac{Z_{n+1}^\nu(z,z^*,\mu) 
Z_{n-1}^\nu(z,z^*,\mu)}
{[Z_{n}^\nu(z, z^*,\mu)]^2}.\hspace*{0.9cm} 
\label{todamu}
\ee
which is obtained by applying the
Sylvester 
identity  to the determinant in (\ref{hankel-form}).
This equation  can be extended to imaginary chemical potential
and the two-point 
correlation function
\cite{Spl03b,Dam05,Dam06}. For imaginary chemical potential 
there is no transition to a Bose condensed 
state and the $n$-dependent part of 
the free energy vanishes after
differentiation with respect to the masses. The nontrivial
result  in the free energy  is $O(n^2)$ which gives the two-point
correlation function.

The replica limit of
the Toda lattice equation  results in 
the spectral density   
\be
\rho(z,z^*,\mu) = \lim_{n\to 0} \frac 1{n \pi}\frac d{dz}\frac d{dz^*}
\log Z_{n}^\nu(z,z^*,\mu)
= \frac 12 z z^*Z_{1}^\nu(z,z^*,\mu) Z_{-1}^\nu(z,z^*,\mu),\hspace*{1cm}
\label{rhomuex}
\ee
which was derived in \cite{Spl03b}.
The fermionic partition function can be obtained by an explicit
evaluation of the integral over  $U(2)$. The result is given by
\be
Z_1^\nu(z,z^*,\mu) = \frac 1\pi e^{2VF^2\mu^2} \int_0^1 d\lambda \lambda e^{-2VF^2\mu^2\lambda^2}
I_\nu(\lambda z \Sigma V) I_\nu(\lambda z^* \Sigma V).
\label{z1}
\ee
The evaluation of the bosonic partition is more complicated. The 
inverse complex conjugated determinants can
only be represented as a Gaussian integral after combining
them into a Hermitian matrix. In order to obtain a convergent integral, 
the Hermitian matrix has to regularized by  a mass $\sim \epsilon$.
This procedure is known as Hermitization \cite{Jan97,Fei97}.
It turns out that the partition function is logarithmically divergent in
$\epsilon$. This divergence is due to a single eigenvalue close to $z$
and is present even if $z$ is
outside  the support of the Dirac spectrum \cite{Spl03b,Spl08}.
 Because of the
Vandermonde determinant, the probability of finding two eigenvalues close
to $z$ does not diverge.
\begin{figure}[t!]
\unitlength1cm
\begin{center}
\begin{picture}(14.2,4.0)
\centerline{\hspace*{-1cm} \epsfig{file=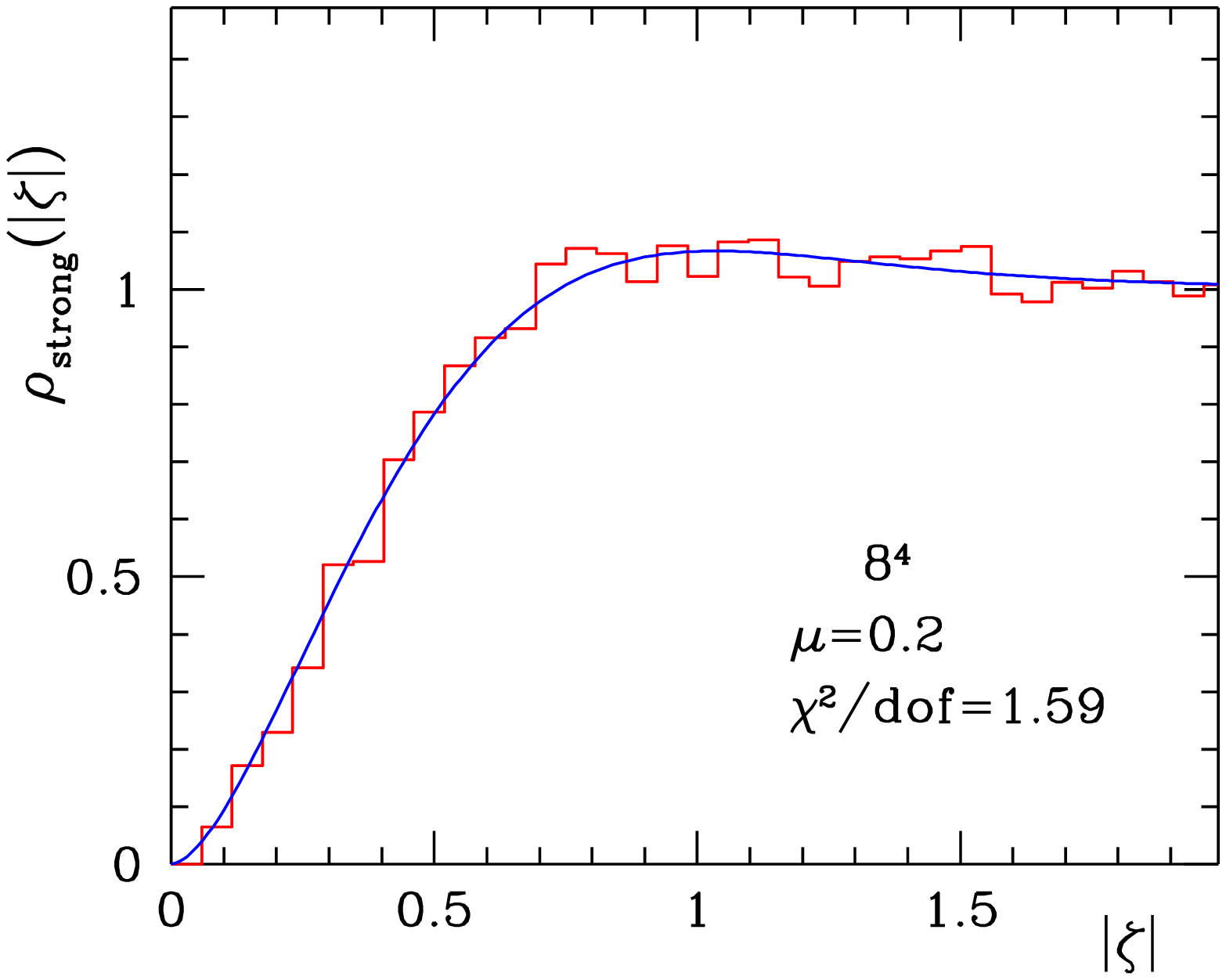,width=5.0cm}\hspace*{1cm}\epsfig{file=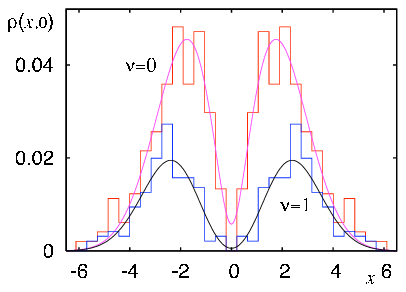,width=5.6cm}}
\end{picture}
\end{center}\vspace*{-0.5cm}
\caption{Left: Radial microscopic spectral density for quenched QCD at 
$\mu \ne 0$ \cite{Wet04}. Right: Spectral density of the overlap Dirac 
operator
as a function of the
distance, $x$, to the imaginary axis \cite{Blo06}.
\label{fig:mmu}}
\end{figure}
The partition function for one pair of conjugate quarks  can also be written
as an integral over Goldstone bosons. 
Instead of an integral over
$U(2)$, using an extension of the Ingham-Siegel integral \cite{Fyo01}, 
we obtain an integral over the noncompact manifold of positive
definite Hermitian matrices $Q$ 
\be
Z_{-1}^{\nu}(z,z^*;\mu) &=& 
\lim_{\epsilon\to0} C_\epsilon
\int \frac{dQ}{{\det}^2 Q} \theta(Q) 
e^{ {\rm Tr} \left[ 
 i \frac{V\Sigma}{2}\zeta^T(Q -I Q^{-1}I )   
-\frac{V}{4}F^2 \mu^2 [Q ,\sigma_3][Q^{-1} ,\sigma_3 ]\right]} ,\nn \\
&& {\rm with} \quad
I = \mat 0 & 1 \\ -1 & 0 \emat, \qquad
\zeta = \mat \epsilon & z \\ z^* & \epsilon \emat.
\label{ZMINQCD}
\ee
The integral over $Q$ can be performed analytically resulting in
\be
Z_{-1}^\nu(z,z^*;\mu) 
& =& 
\frac {C_{-1}e^{-V\mu^2F^2} }{4\mu^2F^2V }
e^{\frac{V\Sigma^2 (y^2-x^2)}{4\mu^2F^2}} K_\nu(\frac{V \Sigma^2(x^2+y^2)}{4\mu^2F^2}).\label{zm1final}
\ee
In Fig. \ref{fig:mmu} we compare the expression for the spectral density to 
quenched lattice simulations \cite{Wet04}. 
The lattice data in the left panel of Fig. \ref{fig:mmu} are in the
strong-nonhermiticity domain where ${\rm Re(z)} \Sigma /2\mu^2 F^2 < 1$ and
$\mu^2 F^2 V \gg 1$. In this domain
the analytical result can be 
simplified to \cite{Spl03b,Ver05}
\be
\rho_{N_f=0, \nu}(z,z^*,u) = \frac 2\pi u^2 zz^* K_\nu(uz^*z)I_\nu(uz^*z),
\ee
with $u$ defined as
\be
u = \frac{\Sigma^2 V}{4 \mu^2F^2}.
\ee
The topological index of the staggered lattice Dirac operator
at relatively strong coupling in the left panel of 
Fig. \ref{fig:mmu} is zero.
Lattice results for nonzero topological charge have been obtained
using the Bloch-Wettig overlap Dirac operator \cite{Blo06} and are compared to the analytical expression (\ref{rhomuex})
in the 
right panel of Fig. \ref{fig:mmu}.

Using superbosonization \cite{Hac95,Bun07,Bas07}, 
the fermionic and bosonic partition function can be combined into a supersymmetric partition  function \cite{Bas07} which can be used to derive the low-energy limit
of the generating function of the QCD Dirac spectrum at $\mu \ne 0$.

\subsection{Spectral Density at $\mu \ne 0$ for QCD with Dynamical Quarks}

\begin{figure}[t!]
\unitlength1cm
\begin{center}
\begin{picture}(14.2,4.0)
\centerline{\epsfig{file=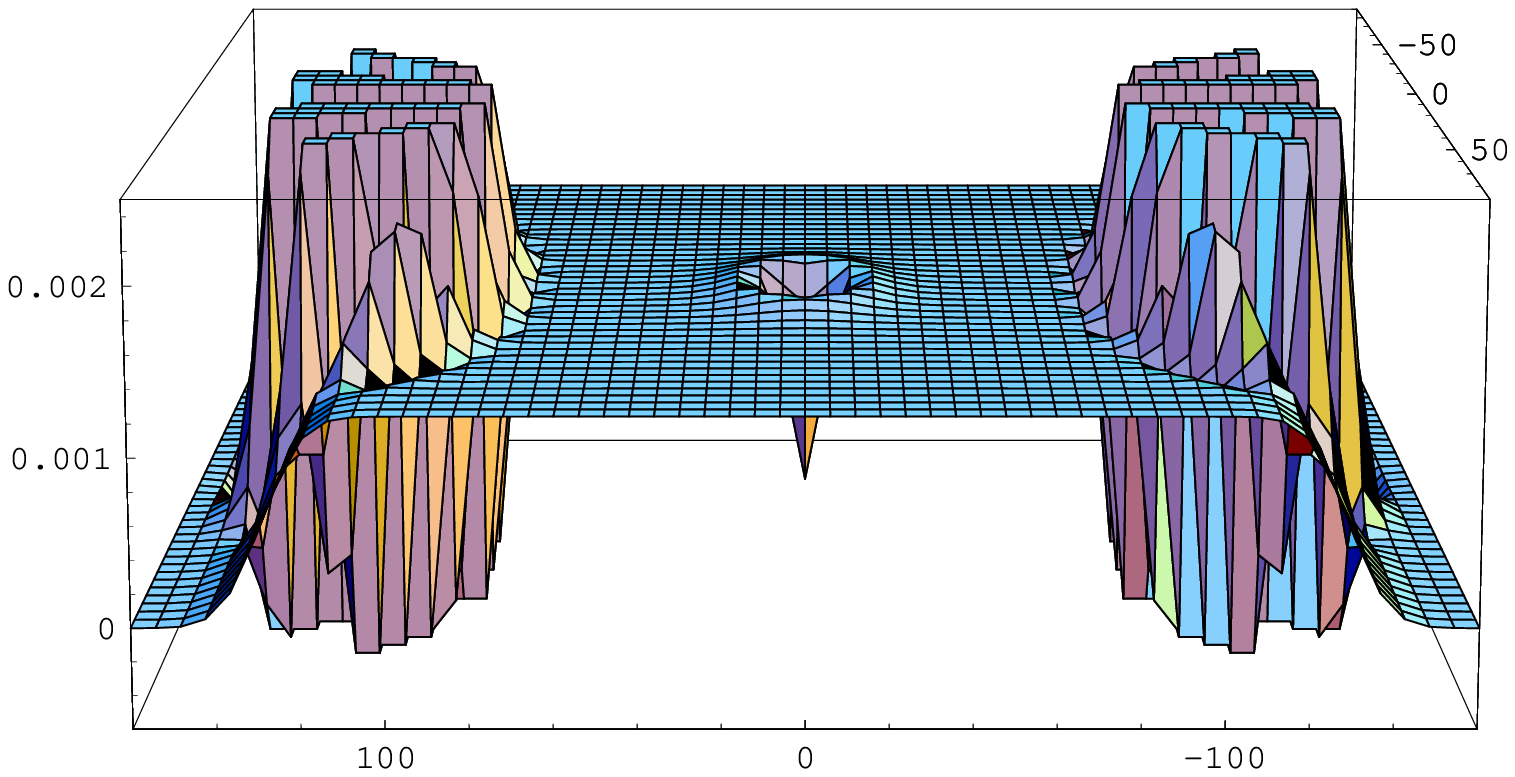,width=7.2cm}\hspace*{1cm}
\epsfig{file=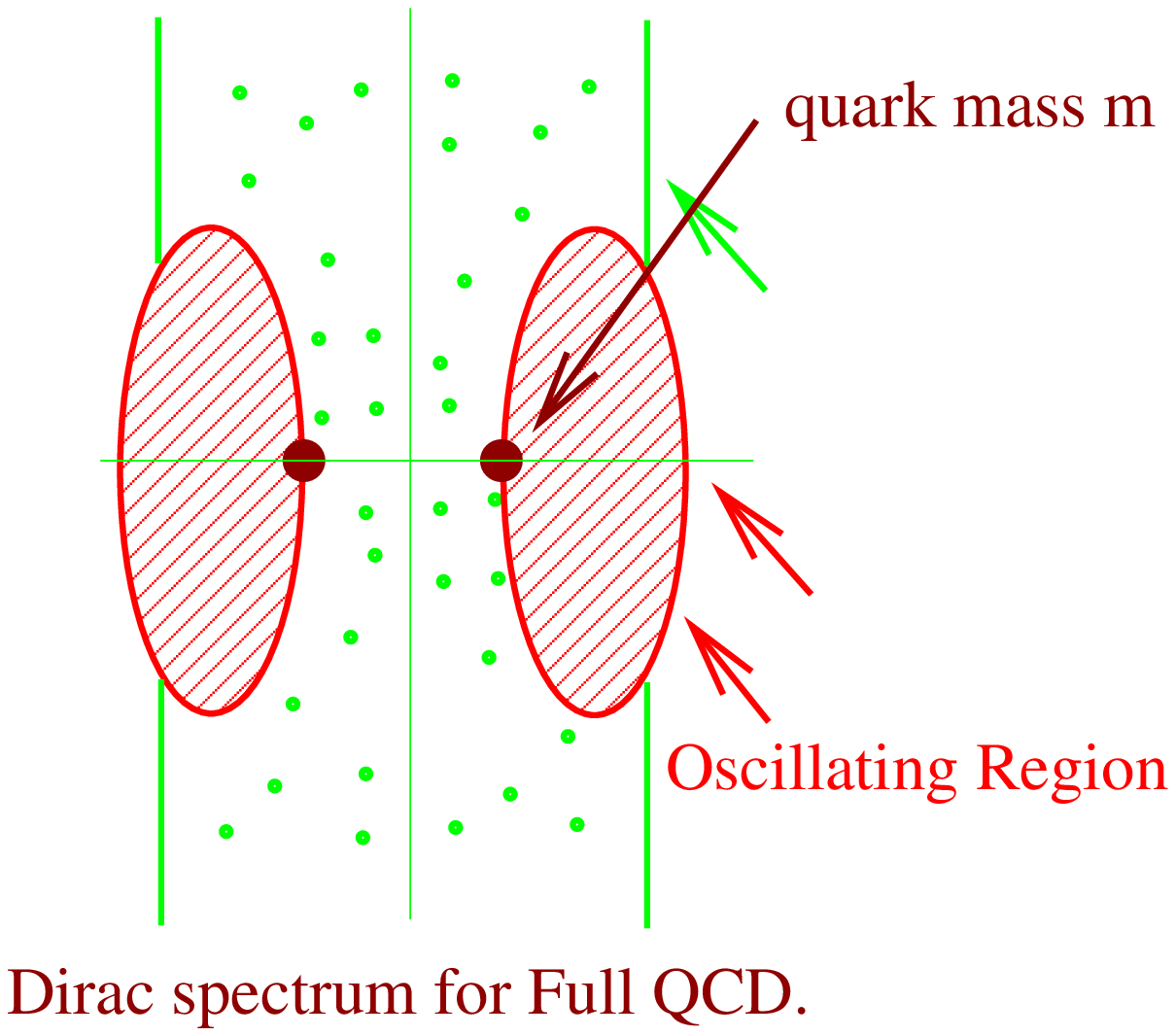,width=4.5cm}}

\put(-1.2,3){\color{green} $\frac {2F^2\mu^2}{\Sigma}$}
\put(-1,1.7){\color{red} $\frac 83 \frac {\mu^2F^2}{\Sigma} -\frac m3$}
\end{picture}
\end{center}
\caption{Left: The real part of the spectral density of the QCD Dirac operator
for one-flavor QCD at $\mu \ne 0$. For better illustration the $z$-axis
has been clipped. Right: The phase diagram of the Dirac spectrum 
\label{fig:nf1dens}}
\end{figure}

Although  the spectral density of the Dirac
operator for  QCD  with
dynamical quarks at $\mu\ne 0$ was first derived using complex
orthogonal polynomials \cite{Osb04}, a simpler expression is obtained from
the Toda lattice equation  \cite{Ake04},
\be
\rho^{\nu}_{N_f}(z,z^*,\mu) \sim  z z^* \prod_{f=1}^{N_f}(m_f^2- z^2)
\frac{Z^{n=-1}(z,z^*,\{ m_f\},\mu) Z^{n=1,N_f}(z,z^*,\{m_f\},\mu)}{Z^{N_f}(\{m_f\})}.
\nn\\[-0.5cm]
\label{rhogen}
\ee
In Fig. \ref{fig:nf1dens} we show a 3d plot of its real
part. In addition to a flat region there is
a strongly oscillating region with oscillations with an amplitude that
increase exponentially with the volume and period that goes like $\sim 1/V$.
This region is absent in the quenched or phase-quenched case and is responsible for the
discontinuity in the chiral condensate.
As we can see from Fig. \ref{fig:nf1dens}, 
 we can distinguish three phases in the Dirac spectrum. 
A phase with no eigenvalues,
a phase with a constant eigenvalue density, and a phase with a strongly oscillating 
eigenvalue density. These phases can be obtained \cite{Osb08b}
by means of a mean field study of a partition function
with masses $m$, $z$ and $z^*$,
similar to the analysis of 3 flavor QCD at nonzero  chemical potentials
 \cite{Kog01}. 
The phase diagram is shown in the right panel of in Fig. \ref{fig:nf1dens}.

As we argued before, in the microscopic domain, the chiral condensate does not depend on the
chemical potential. Both the flat region and oscillating region give rise  to 
a $\mu$ dependent
contribution to the chiral condensate, but the $\mu$-dependence cancels in their sum \cite{Osb05}. This solves the ``Silver Blaze Problem'' \cite{Coh03}
and can be explained \cite{Osb08a}
in terms of orthogonality relations of the complex orthogonal polynomials.

\subsection{The Phase of the Fermion Determinant}

ChRMT can be used to study the complex phase of the 
fermion determinant. 
The average phase factor may be calculated with respect to the quenched,
the phase quenched, or the two-flavor partition function. The phase quenched
average is given by
\be
\langle e^{2i\theta} \rangle_{1+1^*}
\equiv  \left \langle \frac{ \det (D + m+\mu \gamma_0)}
{\det(D^\dagger + m+\mu \gamma_0) } \right \rangle_{1+1^*} = \frac{Z_{1+1}}{Z_{1+1^*}}.
\label{pqph}
\ee
Since $Z_{1+1}$ does not depend on $\mu$,
the latter ratio follows immediately from the expression from $Z_{1+1^*}$ given
in (\ref{zeff}).

The quenched average phase factor 
can be re-written in terms  of
a determinant of complex orthogonal polynomials. In the microscopic domain,
the result is the sum of a polynomial in $\mu^2$ and a part
with an essential singularity at
$\mu = 0$. Therefore it cannot be obtained from analytical continuation
from an imaginary chemical potential which is polynomial in $\mu^2$
  \cite{Dam05,Spl06,Blo08}.

 \subsection {QCD at Imaginary Chemical Potential}
Random matrix models at imaginary chemical potential are obtained by  
replacing $\mu \to i \mu$ in the Dirac operator. Then the Dirac operator 
becomes anti-hermitian with all eigenvalues on the imaginary axis.
There are two  ways of introducing 
an imaginary chemical potential, either as a multiple of the  identity
or as a multiple of a complex random matrix ensemble. Both models have the same
symmetry properties and lead to the same universal partition function 
in the microscopic domain. Spectral correlation functions can be obtained
by means of the Toda lattice equation \cite{Dam06}, or in the
second model, by means
of the method of bi-orthogonal polynomials \cite{Ake08a}.

Parametric correlations of Dirac spectra in the microscopic domain depend
 on two low-energy constants, $ F$ 
and $\Sigma $, which can be extracted 
from correlations of
lattice QCD Dirac spectra \cite{Dam06,DeG07,Ake08a}.

\section{Applications to  Gauge Degrees of Freedom}

The Eguchi-Kawai model \cite{Egu82}  is the lattice Yang-Mills  
partition function with all links
in the same direction  identified. In the large $N_c$ limit
this model is  an integral
over $U(N_c)$-matrices. Although the original hope, that
Wilson loops of Yang-Mills theory are given by this reduced theory 
is incorrect, the model continues to attract a considerable 
amount of attention.  

For $d=4$ the Eguchi-Kawai model cannot be solved analytically, but for $d=2$
it is known as the Brezin-Gross-Witten model \cite{Bre80,Gro80} 
\be
Z = \int _{U \in U(N_c)} dU e^{\frac 1{g^2}{\rm Tr}(U + U^\dagger)}.
\ee
and is identical to 
the zero momentum partition function (\ref{zm}) (see Chapter 17 for a dsicussion of such group integrals).
 In the large $N_c$ limit this model undergoes a third order phase transition
at $g^2N_c = 2$.

In the large $N_c$ limit eigenvalues of Wilson loops can be analyzed by means of RMT methods. 
It was shown in \cite{Dur80} 
that Wilson loops in two dimensions undergo a phase
transition for $N_c \to \infty$ at a critical value of the length of the loop. 
In one phase, the
eigenvalues of the Wilson loop are distributed homogeneously over the unit circle, 
whereas in the
other phase, they are localized at zero. This phase transition has been observed 
in lattice QCD simulations
\cite{Nar06} and has been analyzed in terms of shock solutions of the Burgers equation 
\cite{Bla08,Bla09,Neu08}. It can be studied by analyzing
the eigenvalue distribution of  products of unitary matrices
\cite{Gud03,Loh08}.

\section{Concluding Remarks}

Random Matrix Theory has changed our perspective of the QCD Dirac spectrum. Before the
advent of chiral Random Matrix Theory, the discrete structure of the Dirac spectrum
was viewed as random noise that will go away in the continuum limit. Now we know that
Dirac eigenvalues show intricate correlations that are determined by
chiral random matrix theory with one or two low-energy constants as parameters.
This implies that we can extract the chiral condensate and the pion decay constant
from the distribution of individual eigenvalues.

Chiral Random Matrix Theory primarily applies to the Dirac spectrum, and therefore
we have to distinguish QCD at zero chemical potential, when the Dirac operator
is anti-Hermitian, and QCD at nonzero chemical potential with a nonhermitian Dirac operator. 
In the Hermitian case the statistical properties of the low-lying Dirac 
eigenvalues are completely determined
by the chiral condensate. In the nonhermitian case 
they
are determined by two parameters, the chiral condensate, and
the pion decay constant. Since the 
nonhermitian Dirac spectrum has the geometry of a
strip, the two low-energy constants are determined 
by the eigenvalue density and the
width of the spectrum.

For imaginary chemical potential, the Dirac operator is Hermitian. Although
spectral correlations are completely determined by the chiral condensate, this is
not the case for parametric correlations, which are correlations of eigenvalues
for two different values of the chemical potential. They require both the chiral condensate and the pion decay
constant as input parameters. 

Chiral Random Matrix Theory applies to the low-lying Dirac spectrum. The scale
is set by the momentum dependent terms in the chiral Lagrangian. Physically, it is 
the scale of the quark mass for which the Compton wave length of the Goldstone bosons
is much larger the size of the box. The scale of the chemical potential should
also be well below the inverse box size.

There are two mechanisms to explain confinement in QCD: by condensation of monopoles,
or by the disorder of gauge fields. The success of Random Matrix Theory points
to the second mechanism. It our hope that the work discussed in this
chapter will contribute to the solution of this problem.

{\sc Acknowledgments}:
This work was supported  by U.S. DOE Grant No. DE-FG-88ER40388. Poul Damgaard and
Kim Splittorff are thanked for a critical reading of the manuscript.

\end{document}